\documentclass{pic2012}
\def\runauthor{Jan Ebr for The Pierre Auger Collaboration}
\def\shorttitle{Proton-air cross-section at 57 TeV with Auger}
\begin{document}

\title{Measurement of the proton-air cross-section at sqrt(s) = 57 TeV with the Pierre Auger Observatory 
}

\author{Jan Ebr\footnotemark~~~for The Pierre Auger Collaboration\footnotemark}

\footnotetext{*Institute of Physics of the Academy of Sciences of the Czech Republic, Na Slovance 1999/2, 18221 Prague 8, Czech
Republic}
\footnotetext{$\dagger$Full author list: http://www.auger.org/archive/authors\_2012\_09.html}
\address{Observatorio Pierre Auger, Av. San Mart\'{i}n Norte 304, 5613 Malarg\"{u}e, Argentina
E-mail: ebr@fzu.cz }

\maketitle

\abstracts{ Using measured events from the fluorescence
detector of the Pierre Auger Observatory, an unbiased distribution of the atmospheric slant depths where  showers
reach their maxima has been obtained. Analyzing the tail of this distribution the
proton-air cross-section for particle production at center-of-mass energies per nucleon of 57~TeV
 is determined to be [505$\pm$22(stat)+28$-$36(syst)] mb. Systematic uncertainties in
the analysis arise from the limited knowledge of the primary mass composition, the
need to use shower simulations and the selection of events. For the purpose of
making comparisons with accelerator data we also calculate the inelastic and total
proton-proton cross-sections using an extended Glauber model.
}

\section{The Observatory}The Pierre Auger Observatory is a hybrid detector for ultra-high energy cosmic rays. The surface detector (SD), which detects secondary particles at ground level, consists of 1660 water-Cherenkov detectors spread over 3000 km$^2$ on a triangular grid with 1500 m spacing. The array is overlooked by 27 fluorescence telescopes (the fluorescence detector, FD)  which measure the longitudinal development of extensive air showers (EAS) initiated by the ultra-high energy particles in the atmosphere above the array by detecting the fluorescence and Cherenkov light produced along the shower trajectory as the charged particles cross the atmosphere \cite{1}. The FD can measure events with energies from approximately $10^{17.5}$ eV, while full trigger efficiency in hybrid mode (a fluorescence event in coincidence with at least one tank) is achieved at energies greater than $10^{18}$ eV \cite{energy}.

\section{Observables }
The number of particles in the EAS (or shower size) as a function of the atmospheric slant depth (the amount of atmosphere traversed from its upper edge in g/cm$^2$) is called the shower longitudinal profile. Most of the EAS energy is dissipated through the electromagnetic component. Therefore, the shower size increases until the average energy of the e$^\pm$ in the EAS is about the critical energy of 81 MeV, when the rate of energy loss due to collisions and ionization begins to exceed that due to radiation. Photons have a similar attenuation length due to pair production. The slant depth at which the longitudinal profile of a shower reaches its maximum is called $X_{\mathrm{max}}$ and is one of the primary observables of the fluorescence detector. The distribution of observed values of  $X_{\mathrm{max}}$ is widely used to estimate the mass composition of the primary beam and  it also carries information on the cross-section of the first interaction in the atmosphere \cite{2,3}. 

The differences in $X_{\mathrm{max}}$  between showers of the same primary energy can be due to fluctuations in interactions as well as to different masses of the primary particles. For purely proton primaries, the $X_{\mathrm{max}}$ distribution is a convolution of the fluctuations in the shower development from the point of the first interaction to the shower maximum (which is dependent on details of the hadronic interactions) and the exponential distribution of the depth of the first interaction with mean free path $\lambda_{\mathrm{p-air}} = \langle m_{\mathrm{air}}\rangle/\sigma_{\mathrm{p-air}}$, with the mean target mass of air  $\langle m_{\mathrm{air}}\rangle\approx14.5m_{\mathrm{p}}\approx 24.3\mathrm{\:b\:g/cm^2}$ and the proton-air cross section $\sigma_{\mathrm{p-air}}$ \cite{3}. In a simple  model, heavier nuclei can be considered as the superposition of many nucleons. For an iron nucleus with energy $E$, each nucleon would have an energy $E/56$, and the superposition of 56 lower energy subshowers would reach $X_{\mathrm{max}}$ earlier (higher in the atmosphere). 

Because the shape of the $X_{\mathrm{max}}$ distribution (at least in the range between $10^{18}$ and $10^{18.5}$ eV which we consider) suggests that there is a significant 
fraction of protons in the primary beam (for details on the determination of the primary composition see \cite{4}), we can obtain a proton-enriched sample by using only the deepest observed events. The remaining dependence on both the mass composition and possible photon contamination is evaluated using simulations and included in the systematic uncertainties.  The slant depth range used in the analysis is defined by the slant depths of 20 \% of the most penetrating events, which is a definition that can be easily applied also on simulated events. As the primary observable, we use the exponential slope $\Lambda$ of the tail of the $X_{\mathrm{max}}$ distribution \cite{5}.

\begin{figure}
\centering
\includegraphics[width=95mm]{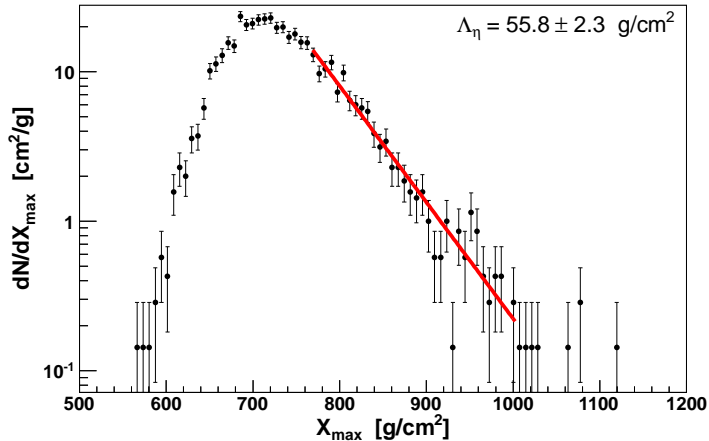}

\caption{The $X_{\mathrm{max}}$  distribution used to obtain the value of $\Lambda$. \label{xdist}}
\end{figure}

\section{Event selection}We use events collected by the fluorescence detector between 1 December 2004 and 20 September 2010 that have a signal in at least one of the SD stations measured in coincidence with the FD.  The geometry for these events is determined with an angular uncertainty of $0.6^{\circ}$. Further we reject events 
\begin{itemize}
\item without reliable measurement of the aerosol optical depth \cite{aerosol},
\item  with excessive cloud coverage (more than 25 \% of the sky),
\item  when $\chi^2$/Ndf is greater than 2.5 when the profile is fitted to a suitable function as this could indicate the presence of residual clouds,
\item with the total uncertainty of $X_{\mathrm{max}}$ greater than 40 g/cm$^2$ 
\item when the angle between the shower and the telescope is smaller than $20^{\circ}$ due to the difficulties of reconstructing their geometry and high fraction of Cherenkov light in such showers,
\item when the reconstructed $X_{\mathrm{max}}$ lies outside the field of view of the fluorescence telescopes.
\end{itemize}
In total, applying these selection criteria yields 11,628 high-quality events \cite{5}.

\section{Fiducial volume selection}

The fluorescence telescopes have a limited field of view in elevation ranging from about  $2^{\circ}$ to  $30^{\circ}$. The geometrical acceptance as well as the limitations related to atmospheric light transmission introduce a bias into the distribution of $X_{\mathrm{max}}$. The bias is even enhanced by demanding that $X_{\mathrm{max}}$ be within the observed profile.  For example, many showers landing close to the FD will have their $X_{\mathrm{max}}$ above the field of view. Some vertical showers, on the other hand, may have their $X_{\mathrm{max}}$ below the ground. In both cases such events will be rejected and the observed $X_{\mathrm{max}}$ distribution will be biased.

To avoid such a bias, we apply the fiducial volume selection. First, we determine from the data the range of values where most (99.8 \%) of the events lie (550--1004 g/cm$^2$). Then, we select only showers with such geometries that each of the showers would be observable for any $X_{\mathrm{max}}$ in the whole range of slant depths (irregardless of the actual $X_{\mathrm{max}}$ of the shower). From this unbiased distribution of 1635 events we find that the 20 \% most deeply penetrating events fall within the range 768 to 1004 g/cm$^2$. Repeating the same procedure for this range, we obtain a set of 3082 events (shown in Fig. \ref{xdist}), of which 783 events contribute to the estimated value of $\Lambda$, yielding  $$\Lambda = 55.8\pm2.3(\mathrm{stat})\pm1.6(\mathrm{syst})\:\mathrm{g/cm^2,}$$
where the systematic error is estimated as the RMS of the distribution of $\Lambda$ calculated using different selection procedures. The average energy of these events is $10^{18.24\pm0.005(\mathrm{stat})}$ eV, corresponding to a center-of-mass energy of $57\pm0.3(\mathrm{stat})$ TeV in proton-proton collisions.

\begin{figure}
\centering
\includegraphics[width=60mm,angle=270]{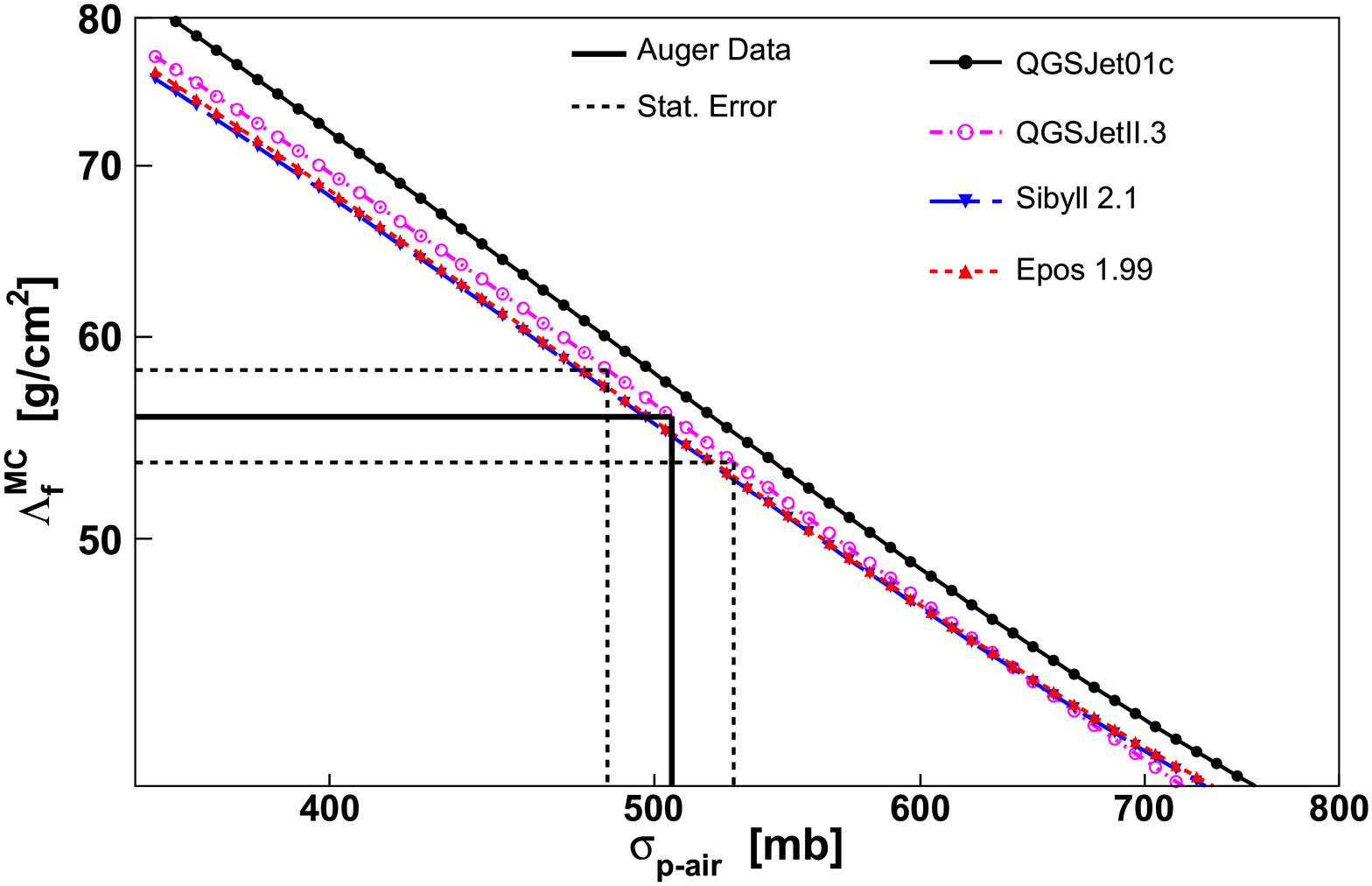}

\caption{Relation between the slope $\Lambda$ in the simulations and the proton-air particle production cross-section \protect\cite{6}.}
\end{figure}

\begin{figure}
\centering
\includegraphics[width=90mm]{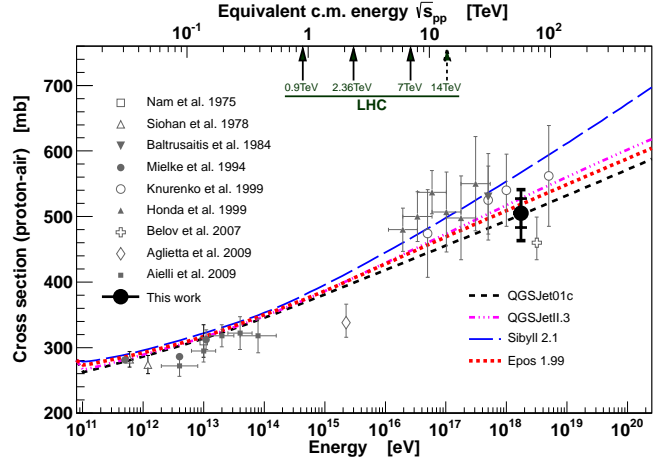}

\caption{Resulting proton-air particle production cross-section compared to other measurements and model predictions. The inner error bars are statistical, while the outer ones include also the systematic uncertainties for a helium fraction of 25 \% and 10 mb for the systematic uncertainty connected with the fraction of photons.}
\end{figure}

\begin{table}
\centering
\caption{Summary of the systematic uncertainties}
\begin{tabular}{l r}
\bf{Description}&\bf{Impact on $\sigma_{\mathrm{p-air}}^{\mathrm{prod}}$} \\
\hline
$\Lambda$ systematics& $\pm15$ mb \\
Hadronic interaction models&$-8, +19$ mb\\
Energy scale&$\pm7$ mb \\
Conversion of  $\Lambda$  to $\sigma_{\mathrm{p-air}}^{\mathrm{prod}}$&$\pm±7$ mb \\
Photons, $< 0.5$ \%&	$< +10$ mb \\
Helium, 10 \% &	$-12$ mb \\
Helium, 25 \% &	$-30$ mb \\
Helium, 50 \% &	$-80$ mb \\
\hline
\bf{Total (25\% helium)}& \bf{$-36$ mb, $+28$ mb}

\end{tabular}

\end{table}

\section{Conversion to cross-section}
We have used the four high-energy hadronic interaction models: QGSJET01 \cite{8}, QGSJETII.3 \cite{9}, SIBYLL 2.1 \cite{10}, and EPOS1.99 \cite{11} to perform Monte Carlo simulations, where we smoothly modify all hadronic cross sections by an energy-dependent factor $f$, given by
$$f\left( E,f_{19}\right)=1+\left(f_{19}-1\right)\frac{\mathrm{ln}\left(E/10^{15}\mathrm{eV}\right)}{\mathrm{ln}\left(10^{19}\mathrm{eV}/10^{15}\mathrm{eV}\right)}$$
($f = 1$ for $ E <10^{15}$ eV) where $E$ denotes the shower energy and $f_{19}$ is the factor by which the cross section is rescaled at $10^{19}$ eV. For each hadronic interaction model, we find a value of $f_{19}$ that reproduces the measured value of $\Lambda$ and use it to evaluate the cross-section at $E = 10^{18.24}$ eV. The proton-air cross sections for particle production obtained by this  procedure deviate from the original model predictions by less than 5 \% in all cases, except for SIBYLL, for which it is 12 \% smaller than the original prediction.

Additional sources of systematic uncertainties are (see Table 1):
\begin{itemize}
\item the systematic uncertainty of 22 \%  in the absolute value of the energy scale of the primary particles \cite{dawson},
\item the dependence of the procedure for retrieving the cross-section from $\Lambda$ on additional parameters such as the energy distribution, energy and $X_{\mathrm{max}}$ resolution,
\item the presence of photons (observational limits on the fraction of photons are $< 0.5$ \%) and the unknown fraction of helium in the primary beam.
\end{itemize}
The resulting value for the proton-air particle production cross-section is 
$$\sigma_{\mathrm{p-air}}^{\mathrm{prod}} = 505 \pm 22(\mathrm{stat})^{+28}_{-36} (\mathrm{syst}) \:\mathrm{mb}$$
at a center-of-mass energy of $57\pm 0.3(\mathrm{stat})\pm6(\mathrm{syst})$ TeV.

\section{Comparison with accelerator data}
We also calculate the inelastic and total proton-proton cross sections using the Glauber model extended by a two-channel implementation of inelastic intermediate states to account for diffraction dissociation \cite{glauber}. This calculation is model-dependent since neither the parameters nor the physical processes involved are known accurately at the cosmic-ray energies. In particular, this applies to the elastic slope parameter $B$, the correlation of $B$ with the cross section, and the cross section for diffractive dissociation. The result for the inelastic proton-proton cross section is 
$$\sigma_{\mathrm{p-p}}^{\mathrm{inel}} = 92 \pm 7(\mathrm{stat})^{+9}_{-11} (\mathrm{syst})\pm 7(\mathrm{Glauber}) \:\mathrm{mb}$$
and the total proton-proton cross section is 
$$\sigma_{\mathrm{p-p}}^{\mathrm{tot}} = 133 \pm 13(\mathrm{stat})^{+17}_{-20} (\mathrm{syst})\pm 16(\mathrm{Glauber}) \:\mathrm{mb}$$
These results agree with extrapolation from data at LHC energies to 57 TeV for a limited set of models. We emphasize that the total theoretical uncertainty of converting the proton-air to a proton-proton cross-section may be larger than estimated here, because there are further possible extensions to the Glauber model to account for a variety of other effects, as discussed in \cite{5}.

\begin{figure}
\centering
\includegraphics[width=80mm]{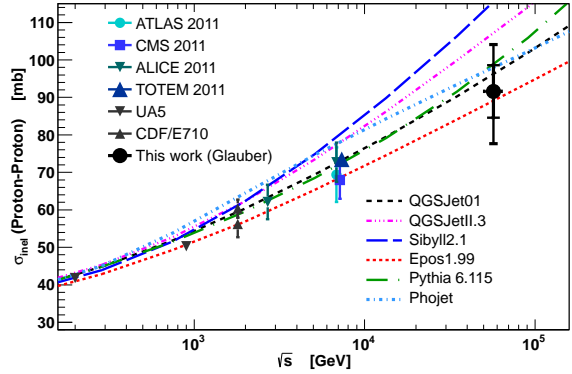}

\caption{Comparison of the estimated inelastic pp cross-section, accelerator data and some model predictions. The inner error bars are statistical, while the
outer ones  include also the systematic uncertainties. For references to the data, see \protect\cite{5}.}
\end{figure}

\section*{Acknowledgements} This contribution is prepared with the support of Ministry of Education, Youth and Sports of the Czech Republic within the project LA08016 and with the support of the Charles University in Prague within the project 119810.

\end{document}